# Do you cite what you tweet? Investigating the relationship between tweeting and citing research articles

Madelaine Hare[1], Geoff Krause[2], Keith MacKnight[3], Timothy D. Bowman[4], Rodrigo Costas[5], Philippe Mongeon[6]

[1] maddie.hare@dal.ca
https://orcid.org/0000-0002-2123-9518
Department of Information Science, Dalhousie University, Canada
*Corresponding author

[2] gkrause@dal.ca
https://orcid.org/0000-0001-7943-5119
Department of Information Science, Dalhousie University, Canada

[3] kt531164@dal.ca
https://orcid.org/0009-0001-0059-1926
Department of Information Science, Dalhousie University, Canada

[4] timothy.d.bowman@wayne.edu
https://orcid.org/0000-0003-0247-4771
Wayne State University, USA

[5] rcostas@cwts.leidenuniv.nl
https://orcid.org/0000-0002-7465-6462
Centre for Science and Technology Studies (CWTS), Leiden University, Netherlands
DSI-NRF Centre of Excellence in Scientometrics and Science, Technology and Innovation Policy (SciSTIP), Stellenbosch University, South Africa

[6] pmongeon@dal.ca
https://orcid.org/0000-0003-1021-059X
Department of Information Science, Dalhousie University, Canada
Centre interuniversitaire de recherche sur la science et la technologie (CIRST), Université du Québec à Montréal, Canada

## Keywords

Altmetrics, Bibliometrics, Indicators, Information Behavior, Social Media, Twitter

## Abstract

The last decade of altmetrics research has demonstrated that altmetrics have a low to moderate correlation with citations, depending on the platform and the discipline, among other factors. Most past studies used academic works as their unit of analysis to determine whether the attention they received on Twitter was a good predictor of academic engagement. Our work revisits the relationship between tweets and citations where the tweet itself is the unit of analysis, and the question is to determine if, at the individual level, the act of tweeting an academic work can shed light on the likelihood of the act of citing that same work. We model this relationship by considering the research activity of the tweeter and its relationship to the tweeted work. Results show that tweeters are more likely to cite works affiliated with their same institution, works published in journals in which they also have published, and works in which they hold authorship. It finds that the older the academic age of a tweeter the less likely they are to cite what they tweet, though there is a positive relationship between citations and the number of works they have published and references they have accumulated over time.

## 1. Introduction

In the early days of altmetric research, much attention was aimed at measuring the "impact"[1] of a publication or set of publications outside of academic research. This was done by counting the number of mentions of a specific work in various types of non-scholarly documents or platforms, like policy documents, Twitter, Wikipedia, etc., which are not the typical document types covered by traditional bibliometric databases like Web of Science and Scopus. Altmetrics did not constitute a fundamental departure from traditional metrics because, like citations, most altmetrics are essentially counts of mentions or references to a scholarly document in other non-scholarly documents. Altmetrics also did not initially break from the traditional focus on peer-reviewed research outputs, which remained the focal units that accumulate these references. However, they did allow for a broader conceptualization and operationalization of the "impact" achieved by scholarly works by capturing engagement with them in the news, in policy documents, in social media, online, etc. Aside from measuring non-scholarly "impact", the second focus of early altmetrics research was the prediction of scholarly "impact". Some early altmetric work (e.g., Eysenbach, 2011) fueled hopes that tweets and other altmetrics, because they tend to accumulate quickly following the publication of a research article (as opposed to citations which can take years to accumulate), could provide an early estimation of the future number of citations the work would receive. Hence, as our literature review will emphasize, much published scholarship sought to determine which, and to what extent, altmetric indicators correlated with citations. However, based on repeated observations that correlation coefficients were weak or moderate at best, the community soon reached the conclusion that most altmetrics are poorly correlated with future citations.

---

[1] We use quotation marks around the word "impact" because despite its abundant use in literature both in and out of the bibliometrics field, the term has also long been criticized for its ambiguity, which in turn makes the different metrics (old and new) imperfect measures of it.

Over the past decade, issues related to the use and interpretation of altmetrics in assessing research quality, measuring the broader impact of research, and information-sharing behaviours have been the subject of scholarly attention (Bornmann, 2015b; Bornmann & Haunschild, 2018; Nuzzolese et al., 2019). Early on, Haustein (2016) summarized three grand challenges for altmetrics: 1) the lack of conceptual frameworks and theoretical foundations to guide the interpretation and understanding of the metrics; 2) the heterogeneity of metrics in terms of platforms, purposes, functions, data sources, indicators; and 3) the quality of altmetric data and their dependencies on platforms, their owners and other stakeholders. These challenges, especially the conceptual ones, point to a need for more research aimed at gaining a deeper understanding of the scope and meaning of altmetrics. As Konkiel (2016) aptly noted, much like citation counts, altmetrics cannot be properly interpreted if used in isolation.

Attempts to interpret altmetrics through novel approaches have occurred in the past decades. Using social media events for the purposes of scholarly metrics requires us to understand the meaning and reason for the activity, which is not easily done. As Haustein (2016) argues, the same events on the same social media platforms can occur for different reasons. For example, a researcher might tweet a publication to promote their own work, to share something relevant to their field, or for the purposes of criticism (Haustein, 2016). Scholars have repeatedly identified the accompanying context of altmetrics as rich sources of data that can assist with this challenge.

Haustein et al. (2016) introduced the notion of describing acts that lead to online events upon which metrics are based. They established a framework that classified these acts into three categories which capture stages and facets of interactions with research objects: access, appraise, and apply (Haustein et al., 2016). Under this framework, a tweet would fall under the appraise category which indicates engagement with the object, and a citation would fall under the apply category, in which a work is used in new knowledge creation. In applying theory to different acts, Haustein et al. (2016) note that tweeting (as an act of appraisal) seems to be relatively free of the Mertonian notion of valuing knowledge claims and scientific merit and is rather the result of other factors such as visibility, social capital, the narrowing of meaning of the publication, or reinforcing one's identity as an expert in a domain. Díaz-Faes et al. (2019) argued for new metrics that shift the focus away from publications and their reception on social media to concentrate on users, their online activity, and their interactions with social media objects. Such interactive perspectives were further conceptualized by Costas et al. (2021) as "heterogenous couplings", understood as the co-occurrence of science and non-science objects, actors and interactions in online and non-scholarly settings. Costas et al. (2021) suggested using network approaches to examine altmetric acts where the focal point is not necessarily the scholarly objects, in addition to expanding to other forms of communication across science and society (e.g. science communication).

Our study contributes to this small but growing body of research by conceptualizing altmetrics as indicators of information behaviour that are better understood when considering the relational characteristics of the actor(s), the work(s)[2] acted upon, and the relationships and interactions between them that exist outside of the context in which the act occurred. Past work in this area includes the studies by Mongeon (2018) and Mongeon et al. (2018), which used social and topical distance to characterize tweets, the work by Díaz-Faes et al. (2019) that characterized researchers based on their activities on Twitter, and the work by Ferreira et al. (2021) that examined authorship, citation, and tweet interactions between researchers and publications at the researcher level. Conducting such studies on a large scale has been facilitated by the recent publication of datasets matching Twitter users with individual researchers (Costas et al., 2020; Mongeon et al., 2023)

Shu and Haustein (2017), unable to infer causation between Twitter visibility and citations received, noted that no direct link has been established at the article level. Our work contributes to this area of scholarship by examining the relationship between tweets and citations from an information behaviour perspective, focussing on tweeting as an *act* involving a specific researcher (the tweeter) and a specific tweeted academic work, and considering the relationship between the tweeter and the tweeted work (Haustein et al., 2016). This constitutes a break from past research that focused on the engagement that works receive on different platforms. Specifically, we will

---

[2] We refer to publications interchangeably with "academic works". This term is used as the OpenAlex database conceptualizes the object "works" as scholarly documents comprising journal articles, books, datasets, and theses, which are present in our dataset.

examine the geographical and socio-topical dimensions of the relationship between the tweeter and the academic work, as well as the individual characteristics of both tweeting authors and the tweeted work. Our objectives are to determine the likelihood that a researcher tweeting an academic work will also cite it, and to determine how factors related to the tweeter (e.g., academic age, research activity, tweeting activity) and the relationship between the tweeter and the tweeted works (e.g., shared field, authors, or geographical location, etc.) may affect this likelihood. Specifically, our study seeks to provide answers to the following research questions:

**RQ1.** To what extent does the act of tweeting an academic work increase (or not) the likelihood of citing it?

**RQ2.** How are the individual characteristics of the Tweeter (academic age and total number of tweets, authored works, distinct references) related to the likelihood of citing the academic work?

**RQ3.** How is the geographical proximity (country and institutional affiliation) of the authors of the academic work related to the Tweeter?

**RQ4.** What is the socio-topical relationship (journal and topic) between the Tweeter's research and the academic work?

An understanding of the connection between tweets, tweeters, and citations can illuminate how research is disseminated through social media and engaged with by scholars; further, it could contribute to how we use altmetrics to interpret and understand tweets and citations as a window into the information behaviour of researchers.

## 2. Literature review

In this section, we review the literature on researchers' use of the social media platform Twitter, the factors that affect engagement with academic works on social media, and the relationship between altmetrics and traditional metrics.

### 2.1. Who are the researchers on Twitter?

The adoption of social media platforms in general, and Twitter specifically, has not been uniform across the board. Yu et al. (2019) found that the identities of scholarly tweeters are diversified: 49% of researchers are university-level faculty members and 38% of them belong to the general public, of which the former are observed to be more correlated with citation counts. Ke et al. (2017) investigated the demographics of scientists on Twitter and found that despite a broad adoption across the disciplinary spectrum, social, computer, and information scientists were over-represented, while mathematical, physical, and life scientists were under-represented. Similarly, Costas et al. (2020) found a strong presence of scholars from the social sciences and humanities, and weaker presences in the physical and applied sciences. They also observed an unequal geographic distribution of scholars on Twitter, with 40% of the scholars in their dataset affiliated with an institution in the United States or the United Kingdom, an over-representation of Australia, Canada, Spain, and the Netherlands, and an underrepresentation of China, Japan, and South Korea. Costas et al. (2020) also observed a correlation between research output and scholars' likelihood to use Twitter, although as they note in their paper, they used the publications to match tweeters and researchers, which means that their matching may be biased towards scholars with more publications. Gender differences in Twitter use by researchers have also been observed, but there is a lack consensus on how much or to what extent. Ke et al. (2017) found the gender ratio in their dataset was less skewed toward men than women than is typical for scholarly studies, Costas et al.

(2020) found that male researchers are slightly more likely to be on Twitter than female researchers across all disciplines, and Zhu et al. (2019) observed no significant gender differences in Twitter use in the health sciences.

## 2.2. Why and how researchers use Twitter

There are many reasons for researchers to use Twitter, subsequently informing the specific ways in which they will use the platform. Researchers use Twitter for networking purposes to connect with other academic and non-academic users, share information and resources, converse, stay up to date, manage their public identify, and promote their work (Adie, 2013; Holmberg et al., 2014; Jordan & Weller, 2018; Robinson-Garcia et al., 2018; Singh, 2020). Holmberg and Thelwall (2014) found that most links researchers tweeted led to science blogs, news sites, and magazines rather than to peer-reviewed articles, suggesting scientists use the platform to popularize science. The centralized nature of Twitter makes it a single, practical, and diverse platform useful to receive and to disseminate information related to research and research events (Bonetta, 2009; Webb, 2016). It is also used as a "back channel" for conferences, conference attendees or non-attendees alike (Singh, 2020). Some studies have highlighted the fact that researchers may not necessarily use Twitter solely in this capacity, but that researchers use it for professional and personal purposes alike (Bowman, 2015; Ke et al., 2017). These types of or motivations for Twitter use are enabled by the platform's affordances such as tweeting, including links in tweets, following other users, retweeting or liking their posts, etc.

Holmberg & Thelwall (2014) assert that the discipline should always be considered when analyzing scholars on Twitter. Their investigation of Twitter use by researchers in different disciplines showed that biochemists retweeted the most, that economists shared the most links, and that the use of Twitter for scholarly communication was marginal in disciplines such as economics and sociology as compared to other disciplines like biochemistry, astrophysics, and digital humanities. Other studies focused on specific disciplines investigated the use of Twitter by instructors and students in education (Veletsianos & Kimmons, 2016), astrophysicists (Holmberg et al., 2014), physics (Webb, 2016) and biomedicine (Haustein et al., 2014). Sugimoto et al. (2017) noted a lack of consensus in the literature regarding disciplinary differences in social media use, with findings varying based on population and field delineation.

## 2.3. Engagement with scholarly literature on Twitter and its determinants

The literature has addressed several factors potentially affecting engagement with scholarly literature on Twitter and the degree to which it correlates with citation counts (Haustein et al., 2015). Zhang and Wang (2018) observed that one of the main factors which contributed to a paper's popularity with the public on Twitter was whether it was tweeted close to the date of publication. Similarly, Costas et al. (2014) found that recent publications are more likely to result in altmetric activity. Didegah et al. (2018) observed a negative correlation between journal impact factor (JIF) and tweet counts, and that funded articles received more tweets than non-funded research. There is also some evidence that high altmetric engagement is influenced by whether papers are tweeted by the official social media account of the journal in which they are published (Zhang & Wang, 2018). Engagement with academic work on Twitter is also influenced by its discipline, which could be due to different levels of public interest in particular disciplines on topics, but also to differences in behaviors on Twitter that may be driven by disciplinary norms. Ortega (2018) found that articles placed in a "general" category received the most engagement, and that Health Science and Social Science articles received the least, while Costas et al. (2014) found that engagement is highest amongst the social sciences, humanities, and the medical and life sciences (Costas et al., 2014). Twitter engagement with papers also varies by country. A study by Alperin (2015) found that about 6% of papers from Latin America were tweeted at least once, a much lower engagement than was reported in previous studies. Alperin hypothesized that this could be reflective of the Latin American articles from the dataset having a lower level of usage overall, that social media usage by academics in Latin America is lower than in other studied geographic contexts, or that there exists a different culture of research sharing on Twitter in Latin America. Shu et al. (2018) found that only 29.3% of the citations of tweeted Chinese papers appeared in other Chinese papers, as opposed to 47.0% for non-tweeted papers, which could suggest that tweeted papers received more international attention, or that a self-selection bias makes papers more likely to be tweeted if they have an international appeal.

Paul-Hus et al. (2015) found increased gender parity with Twitter counts compared to citations. Dehdarirad (2020) found a small citation advantage for female-authored papers as compared to male-authored papers with the same number of early tweets. In contrast, a large-scale study by Vásárhelyi et al. (2021) found that online science dissemination is male dominated and that even in areas with greater female representation, women receive fewer online mentions than male authors. Zhang and Wang (2018) examined the relationship between high social media impact using tweets and high citation counts for biology papers and found that highly tweeted articles were mostly tweeted by the public, while highly cited articles were mostly tweeted by scientists and had little traction with the public. Bornmann (2015b) found that papers tagged with "good for teaching" saw heightened engagement on Twitter and Facebook, suggesting that such papers tend to reach a broader audience than other works. Bornmann (2016) and Sud and Thelwall (2014) commented on limitations of correlation in altmetrics-related work, and called for more analyses considering the content of the tweets, which could shed a better light on what tweet counts indeed reflect about research impact (Bornmann, 2016). Some studies have taken the content of the tweets into account by using sentiment analysis (Friedrich et al., 2015; Hassan et al., 2021). Mathematics, computer science, the life and earth sciences, and the social sciences and humanities were found to have the most positive sentiments, while the physical sciences and engineering had the most negative sentiments (Hassan et al., 2021). Bornmann and Haunschild (2018) found citations (and Mendeley reader counts) to be better predictors of scientific quality (as evaluated by peer assessment) than tweets, noting that Mendeley reader counts may be a more reliable than tweets for research evaluation (Bornmann & Haunschild, 2018).

## 2.4. Correlation between altmetrics and citation counts

The wide adoption of Twitter by scholars acted as an impetus for research examining altmetric activity as early indicators of scholarly impact, or of engagement with wider audiences (Sugimoto et al., 2017). Concerted efforts have been made in the scientometrics community to correlate altmetric indicators and citations. The first of these studies, by Eysenbach (2011), found that there were moderate to significant correlations between tweets and citations, and that highly tweeted articles were 11 times more likely to be highly cited than less tweeted articles. It is now generally accepted that there exists a positive (though usually weak to moderate) relationship between tweets and citations (Bornmann, 2015a; Costas et al., 2014; Thelwall et al., 2013). In comparing tweeted papers and non-tweeted papers from the same journal and year of publication, Shu and Haustein (2017) found that tweeted papers received 30% more citations on average than non-tweeted papers. Shu et al. (2018) found average citation rates for Chinese papers to be 50% higher for tweeted papers than non-tweeted papers. Thelwall et al. (2013) found strong evidence for an association between six different altmetric indicators (including tweets) and citation counts, though their study did not provide effect sizes. De Winter (2015) found a minimal relationship between tweet counts and citation counts, concluding that Twitter activity is largely independent of the citation behaviours occurring in the scholarly research system and that high-impact publications may accrue a large number of citations even after Twitter activity has dissipated (de Winter, 2015).

## 2.5. Relationship between research and tweeting practices at the individual level

The growing research on the relationship between citations and altmetrics has therefore observed positive correlations, but the relationship between research and tweeting practices at the individual level is still being investigated. Díaz-Faes et al. (2019) looked at how Twitter users' overall activity related to science communication and considered their specific interactions with research objects. They characterized different social media communities of attention around scientific activities and interactions with scientific results. They advocated for taking the broader perspective of "social media studies of science," which characterize forms of "heterogenous couplings", a concept introduced by Costas et al. (2021), between social media and research objects (Díaz-Faes et al., 2019). This advances social media metrics towards interactive and network perspectives in which metrics are distinguished by those related to research objects and those related to social media users (Díaz-Faes et al., 2019). Ferreira et al. (2021) analyzed Web of Science authors identified on Twitter, comparing tweeted works with academic activities such as citations, self-citations, and authorship at the author level. They also examined the similarity of topics of the publications tweeted, cited, and authored (Ferreira et al., 2021). The productivity and impact of researchers on Twitter bore no correlation to their popularity on the platform and they tended to tweet about topics in close relation to publications they authored and cited. Mongeon (2018) proposed a model based on the social relationship between the content creator and the user and topical distance between the content and the user's main topics of interest to link the content of the objects of social media acts to the social and informational universes of users who perform the acts. His analysis of a set of tweeted JASIST publications meaningfully differentiated between types of heterogenous social media acts as well as characterized the attention specific content received to distinguish different types of social media uses and users. Mongeon et al. (2018) built on this approach by looking at the social and topical distance between tweeted information science papers and their tweeters to examine the relationship between tweets and citations. They showed that researchers were more likely to cite what they tweet when the tweeted paper was related to the tweeter's social network and research topic.

Our study advances approaches to social media metrics by looking at different "heterogenous couplings" as described by Costas et al (2021). Metrics relating to both social media users and tweeted academic works are coupled to link altmetrics to traditional metrics based on different factors relating to the tweeter as a social media user, the tweeter as a researcher, and their interaction with academic works online. We follow Ferreira et al. (2021) by taking into account the characteristics of the tweeter as both a social media user and researcher to examine the relationship between works they tweet and works they cite. We build on Mongeon's (2018) suggestion to incorporate other concepts and dimensions into a model that observes the social and topical dimensions of users and the academic work they engage with. Further, Mongeon et al. (2018) demonstrated that it is possible to distinguish between potentially different tweeting motivations by considering the publication profile of tweeting authors. In looking at the relationship between the publication behaviours and characteristics of tweeting researchers, our study seeks to link tweets to citations by accounting for how these factors influence tweeting motivations. Our study thus employs an interactive approach advocated for by other altmetric scholars to establish a connection between tweets, tweeters, and citations. It diverges from an engagement-only approach that relies on altmetric counts, and instead examines different dimensions of the tweeter-tweeted work relationship (geographical, socio-topical, individual characteristics of the tweeting authors and academic works). By accepting and accounting for heterogeneric couplings of different factors, we provide a more holistic survey of the relationship between researchers as social media users and the work they engage with in online environments and in their research, positioning altmetric activity as information behaviour in the processes of knowledge production and dissemination.

We hypothesize that the topical proximity of the tweeted work to the tweeter's research, as well as the social and geographical proximity (which may overlap with topical proximity), will positively affect the likelihood of a future citation. We also hypothesize that the number of works (and references made) by a researcher will positively be associated to the likelihood of a citation to the tweeted work; it seems plausible that prolific researchers will be statistically more likely to cite tweeted works than those who publish less. Conversely, we hypothesize that the number of tweeted works will negatively be associated to the likelihood of a citation to the tweeted work. Researchers who tweet at high rates are most likely engaging with a high volume of diverse material and content perhaps more superficially or with passive interest, resulting in a lower likelihood that they will cite what they tweet. Finally, we posit that academic age will affect the likelihood of citations as research and publishing skills and practices mature as one's career progresses, although it is not clear if this effect will be positive or negative.

# 3. Data and methods

## 3.1. Data collection

Individual tweets containing references to academic works were obtained from a data dump of Crossref Event Data, circa January 2023[3], which contains a set of more than 81 million tweets, starting in 2017, linked to DOIs of academic works, along with the tweets' metadata. Over 27 million of these occurred in the 2017-2019 timeframe, which was limited to allow time for citations to accrue. Tweets were cross-referenced, via the Twitter handle, to a dataset of handles of known scholars on Twitter, produced by Mongeon et al. (2023); this dataset contains Twitter handles matched to OpenAlex Author IDs of researchers using various combinations of the author names and the Twitter user name or handle. Because some of the matching methods used by Mongeon et al. (2023) are less precise than others, these matches are manually validated in the dataset. Our paper uses 403,710 matches from this data set, which excludes the matches manually flagged as false positives by Mongeon et al. (2023). Tweeters in this set were linked to just over 7 million unique tweets in the timeframe being examined. Both the tweeters and tweeted works (via DOIs) are linked to a mirror of OpenAlex data (Priem et al., 2022), circa May 2022, stored in a PostgreSQL database hosted by the Maritime Institute of Science, Technology and Society (MISTS), in which the OpenAlex venues have been assigned to one or more domain, field, and subfield from the Science Metrix journal classification (Archambault et al., 2011). The field classification of OpenAlex venues[4] is done in two steps. First, we directly match the venue based on the ISSN or name, and then assign the remaining venues to the most cited discipline based on the works cited by the works published by the venue. Overall, 43,318 OpenAlex venues are assigned to at least one discipline (28,904 through direct match, and 14,414 through the cited works). By combining these datasets, we obtain tables in which each observation is a tweeted paper and includes relevant metadata about the tweet, the Twitter user, and their publication record, as well as the tweeted publications and their authors. Approximately 6.4 million tweets made by researchers in our dataset were linked to just over 1 million distinct DOIs found in the OpenAlex works table.

---

[3] Due to changes in the Twitter APIs and the agreement between Crossref and Twitter, starting in February 2023, tweet-related event data, including historical data, is no longer available through the Crossref Event Data APIs (Crossref, n.d.).
[4] Subsequent versions of the OpenAlex schema replace the term "venue" with "source".

## 3.2. Data processing

For each tweet, the tweeter's OpenAlex Author ID was used to retrieve authorship records for all works by that author, using custom SQL queries to the database, producing a list of works authored, journals in which these were published, and linked institutional affiliations. These authorship records were then used to obtain a list of co-authors for the tweeter, the OpenAlex domains/fields/subfields, and the countries in which the institutions are located. The "article-level classification" value for journal classifications is omitted from the list for comparison. The authorship records of the tweeter were then compared to the same information for the authors of the tweeted paper.

Academic age of the tweeter was calculated by subtracting the earliest publication year of any academic works they have authored from the year in which the tweet itself was produced. This value will therefore differ between tweets made by the same tweeter in different years. A negative age may result when the tweeter's first publication occurs after the tweet; this is a valid outcome of the calculation, and such values are included in our results. Although OpenAlex author disambiguation more typically results in authors' work being split across multiple identifiers, the reverse situation has been known to occur, where disparate individuals' authorships are combined into a single identifier. In some cases, this has resulted in academic ages exceeding 100 years;

analyses involving academic age will exclude any observations where this value exceeds 60 years (9480 tweets).

Counts of tweeters' publications, distinct references, and tweeted links to academic works were calculated through database queries to count the relevant records in the OpenAlex and Crossref datasets.

In order to allow for valid comparisons, tweets of academic works lacking journal classifications (339,107) or author-institution links (1,033,589) were omitted. Likewise, tweets by authors lacking any publication or reference data were omitted (297,091). Tweets may be affected by multiple exclusion criteria. The final analyzed dataset totalled 5,307,769 tweets made between 2017 and 2019.

## 3.3. Analysis

This analysis uses several concepts to investigate the relationship between citation behaviours, the tweeter, and their published work. These are operationalized as indicators, reported in Table 1 and elaborated on below.

| Dimension | Indicator | Description |
|---|---|---|
| **Geographical** | same_country | The tweeting author is affiliated to the same country as at least one of the authors of the tweeted publication. |
| | same_institution | The tweeting author is affiliated to the same institution as at least one of the authors of the tweeted publication. |
| **Socio-topical** | same_domain | Tweeting author has at least one publication in the same domain as the tweeted publication. |
| | same_field | The tweeting author has at least one publication in the same field as the tweeted publication. |
| | same_subfield | The tweeting author has at least one publication in the same subfield as the tweeted publication. |
| | same_journal | The tweeting author has at least one publication in the same journal as the tweeted publication. |
| | co_authorship | The tweeting author was a co-author on another work with one or more authors of the tweeted publication. |
| | self_tweet | The tweeting author is an author of the tweeted work. |
| **Individual** | academic_age | The earliest year of publication for the tweeting author subtracted by the year of the tweet. |
| | n_tweeted_works | The total number of tweeted works by a tweeting author. |
| | n_works | The total number of academic works of a tweeting author. |
| | n_references | The total number of distinct references a tweeting author cited cumulatively in their works. |

Table 1. Dimensions and their indicators

The descriptive figures in our results produced using the geographical and socio-topical variables are treated as mutually exclusive within each category; that is, tweets of works by authors at the same institution are not also counted towards the same country, and tweets by authors of a work are not also counted towards co-author or journal matches. For example, if a researcher is affiliated with University of Toronto and cites a work affiliated with that university, it will count towards the same institution, but not the same country. If a researcher from University of Toronto cites a paper from a different university within Canada, it will count towards the same country. Similarly, self-tweets are distinguished from co-authorship in that if a researcher tweets a work they wrote with a co-author, this counts towards a self-tweet, but not a co-authorship. This is done to prevent a conflation of distinct variables. The noted variables are mutually exclusive, but not from other variables.

### 3.3.1. *Statistical model*

Our goal is to illuminate the likelihood of a citation to the tweeted work by the Twitter user, which is recorded in our dependent variable named *cited*. Because several factors are likely to affect the probability that a tweeted work will also be cited, our model includes variables that capture the research activity of the tweeter and its relationship to the tweeted work. More specifically, we consider the number of distinct publications cited by the Twitter users, their geographical distance (country and the institutional affiliation) from the authors, and the topical distance (journal and topic) between the user's own research and the tweeted work.

# 4. Results

## 4.1. Descriptive analysis

Of the 5,307,769 tweets containing links to journal articles, 768,710 corresponded to citations in works authored by the same Twitter user, a rate of 14.5%. Table 2 shows the ranges and descriptive statistics for variables relating to the individual tweeters/authors, both for the entire set of tweets and those corresponding to citations, on a per-tweet basis. The ranges of variables for all tweets and those corresponding to citations are the same; averages for academic age, number of published works, and distinct references are all higher for tweets linked to works cited by the tweeter, while averages of works tweeted is lower amongst tweets that are tied to citations.

| | **All Tweets** | | | | | |
| **Variable** | **Count** | **Min** | **Median** | **Mean** | **Max** | **SD** |
| Academic Age | 5,307,769 | -5 | 10 | 11.73 | 60 | 9.79 |
| Published Works | 5,307,769 | 1 | 31 | 65.06 | 4,008 | 105.65 |
| References | 5,307,769 | 1 | 880 | 1,644.35 | 69,072 | 2,377.74 |
| Tweeted Works | 5,307,769 | 1 | 106 | 319.65 | 7,256 | 685.17 |

| | **Tweets of Cited Works** | | | | | |
| **Variable** | **Count** | **Min** | **Median** | **Mean** | **Max** | **SD** |
| Academic Age | 768,710 | -5 | 11 | 12.65 | 60 | 9.54 |
| Published Works | 768,710 | 1 | 50 | 91.95 | 4008 | 131.61 |
| References | 768,710 | 1 | 1,379 | 2,271.46 | 69,072 | 2,896.18 |
| Tweeted Works | 768,710 | 1 | 55 | 156.55 | 7,256 | 348.28 |

Table 2. Descriptive statistics of individual dimensions

### 4.1.1. Geographical dimensions

The results of our analysis show the relationship between citation rates and country and institution. Figure 1 shows that 6.0% of tweeted works within our dataset, which were created in the same country as the tweeter, were cited by that tweeter, while 37.7% of works from the same institution as the tweeter were cited. 4.3% of the tweeted works with no identified geographical tie between the tweeter and the tweeted work are cited by the tweeter. As indicated in Figure 1, authors are more likely to cite works they tweet if the work was affiliated with the same institution of the tweeter. This likelihood is reduced considerably if the work is affiliated with the same country of the tweeter, though this affiliation does positively increase their likelihood to cite the tweeted publication. Works from the same institution as the tweeter may also have a degree of topical proximity to the work of the tweeter, affecting their likelihood of being cited.

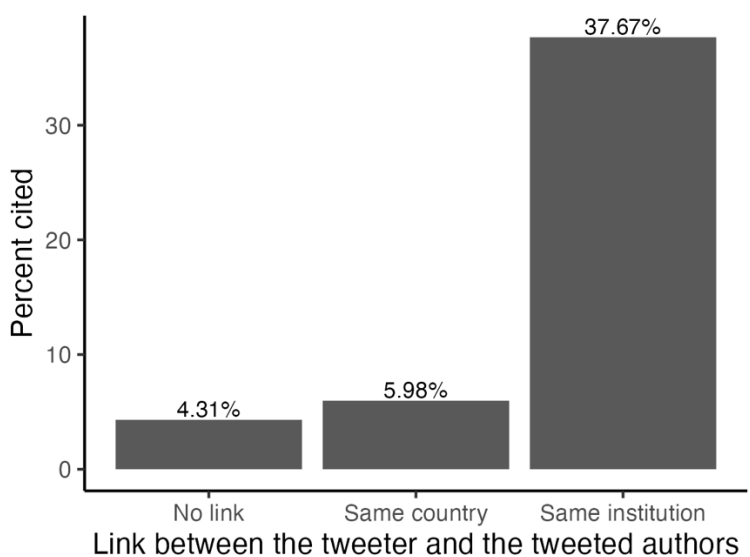

Figure 1. Citation rates by geographic affiliation

## 4.1.2. Socio-topical dimensions

Figure 2 shows the relationship between various socio-topical dimensions and cited works. Our results indicate that 55.89% of cited academic works were a tweeter's own work, meaning a work is more likely to be cited if it was written by the tweeter. Similarly, 22.77% of cited works featured the tweeter as a co-author, meaning a tweet is more likely to result in a citation if the tweeter was a co-author of one or more authors on the tweeted work. 5.94% of cited works were in a journal a tweeter has previously published in, meaning if a tweeting author has at least one publication in the same journal as the tweeted work, this also positively impacts the likelihood of a citation. This may be related to the factor of topical proximity, as academic journals cited works are published in are likely to contain works with similar topics as the tweeter. Further, sub-fields have a small influence on whether a work will be cited (3.9% of cited works), whereas same field (2.22%), domain (1.9%), and those with no link (2.27%) possess relatively equal, but minimal to no relationship to citations. Our results, therefore, indicate that if the topic or discipline of the research object is the same as that of the tweeter it is more likely to be cited, instantiated by the greatest socio-topical influence on whether an academic work is likely to be cited being whether that is authored or co-authored by a tweeter.

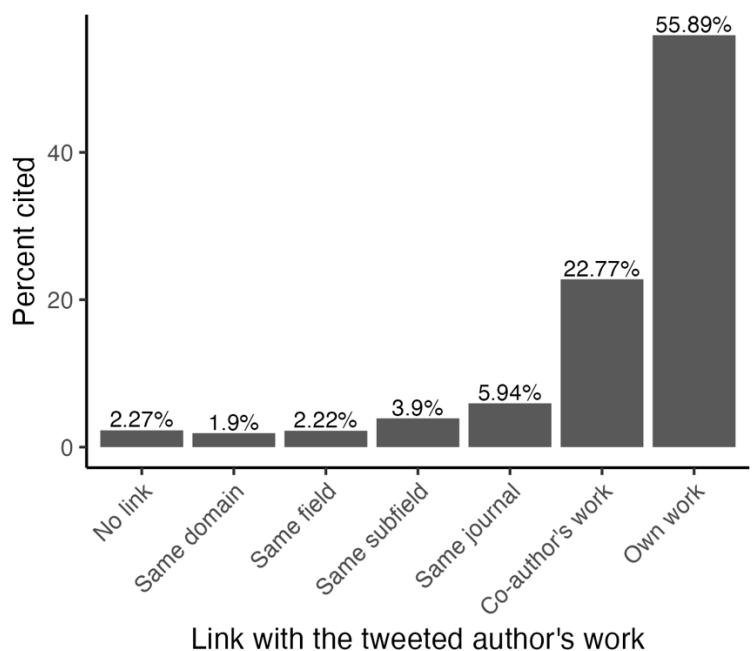

Figure 2. Citation rates by socio-topical characteristics

### 4.1.3. Individual dimensions

Figure 3 shows that the academic age of the tweeter in relation to cited tweeted works. Our results indicate that the likelihood of citing a tweeted work increases quickly in the first years of the academic career, peaks around the tenth year and then plateaus. Beyond 25 years of academic age, the data shows a subtle decline but increased variability in the data.

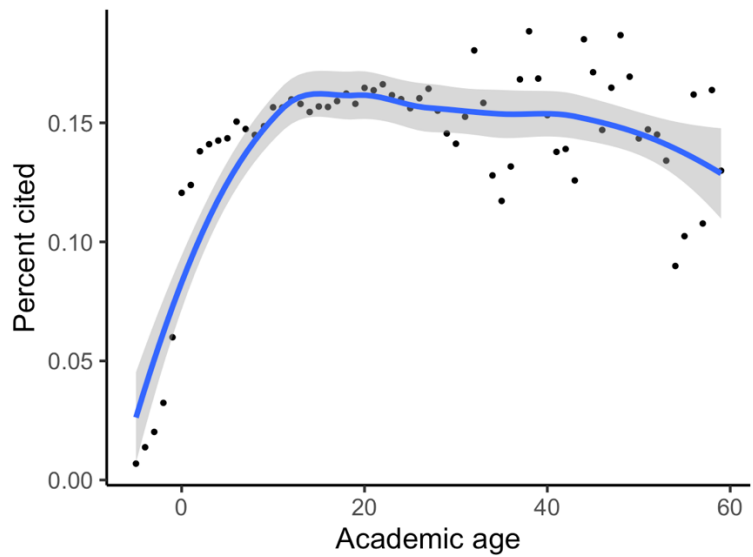

Figure 3. Percent tweeted cited papers rates by academic age

Figure 4 shows a negative correlation between the total number of a researcher's tweeted works and the rate of citation. Authors are less likely to cite what they tweet if they are highly active tweeters of scholarly works.

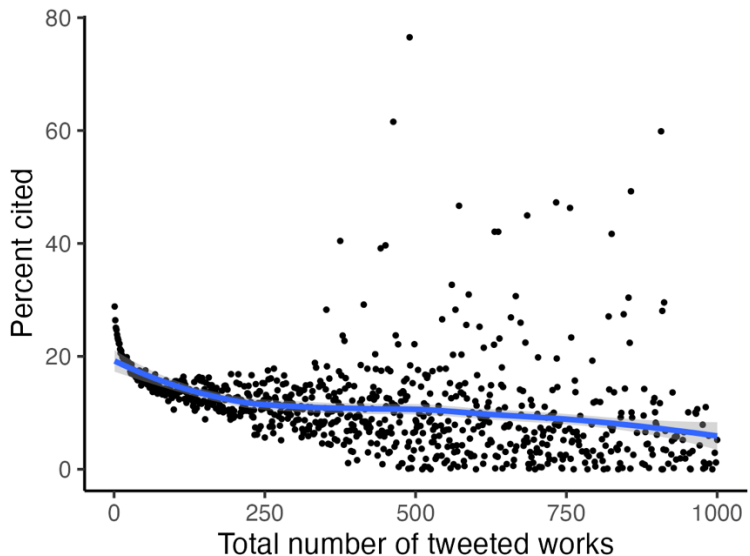

Figure 4. Citation rates by total number of tweeted works

Our analysis finds that the individual characteristics of the tweeter have a relationship with whether a tweeted work will also be cited. Figure 5 shows that the total number of works a tweeter has published in their academic career has a weak but positive correlation with their likelihood of citing tweeted works. A stronger correlation is evident in the first 100 works, indicating that researchers who are more prolific are more likely to cite what they tweet, but this tapers off around 250 publications. Further, the cumulative number of distinct references a tweeter has made also has a positive correlation with their likelihood to cite what they tweet.

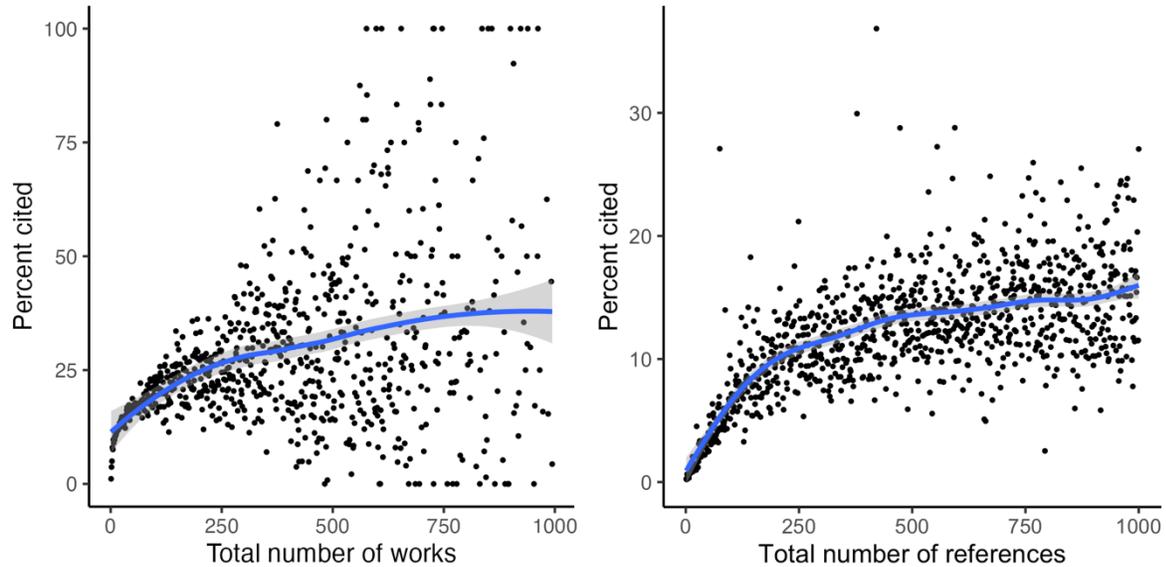

Figure 5. Citation rates by total number of works (left) and distinct cited references (right)

## 4.2. Logistic regression model

We generated a binomial logistic regression model using a random sample of 10,000 tweets from the cleaned dataset to observe the relationship of the dependent, dichotomous variable *cited* and to the independent variables identified in the methods section. These include the categorical variables representing the geographical and socio-topical relationships between the tweeter and the tweeted work, and discrete quantitative variables related to the tweeter's publication and tweeting history.

| Variable | Category | Coeff. | Std. Err. | z value | Pr | Odds ratio | 2.5% | 97.5% |
|---|---|---|---|---|---|---|---|---|
| (Intercept) | -- | -3.583 | 0.170 | -21.085 | 0.000 | 0.028 | 0.020 | 0.039 |
| Country | Geographic | 0.089 | 0.113 | 0.792 | 0.428 | 1.094 | 0.877 | 1.365 |
| Institution | Geographic | 0.012 | 0.137 | 0.084 | 0.933 | 1.012 | 0.773 | 1.323 |
| **Self-Tweet** | Socio-Topical | 3.725 | 0.197 | 18.932 | **0.000** | 41.455 | 28.470 | 61.632 |
| **Coauthor** | Socio-Topical | 2.365 | 0.180 | 13.145 | **0.000** | 10.647 | 7.561 | 15.331 |
| **Journal** | Socio-Topical | 0.818 | 0.187 | 4.371 | **0.000** | 2.265 | 1.582 | 3.301 |
| Subfield | Socio-Topical | 0.281 | 0.194 | 1.452 | 0.147 | 1.325 | 0.912 | 1.953 |
| Field | Socio-Topical | -0.321 | 0.271 | -1.186 | 0.236 | 0.726 | 0.419 | 1.218 |
| Domain | Socio-Topical | -0.343 | 0.356 | -0.962 | 0.336 | 0.710 | 0.334 | 1.370 |
| Papers | Individual | -0.002 | 0.001 | -2.894 | 0.004 | 0.998 | 0.997 | 0.999 |
| **References** | Individual | 0.000 | 0.000 | 5.904 | **0.000** | 1.000 | 1.000 | 1.000 |
| **Tweeted Papers** | Individual | -0.000 | 0.000 | -3.574 | **0.000** | 0.100 | 0.999 | 0.100 |
| Academic Age | Individual | -0.006 | 0.004 | -1.479 | 0.139 | 0.994 | 0.98578 | 1.00196 |

Table 3. Coefficients of the logistic regression model

Of the variables under examination, five were found to have a significant relationship to the dependent *cited* variable; 3 of these relate to socio-topical dimensions (self-tweet, coauthor, journal), and 2 to individual ones (references, tweeted works). In the socio-topical variables, self-tweeted works, works authored by the tweeter's co-authors, and works published in the same journal as the tweeters were found, in decreasing order of strength, to have a positive correlation to whether or not the tweeter would cite the work. Of the individual characteristics of the tweeting author, the number of unique references cited by the tweeter in their own publications was found to have a positive correlation to tweeted works being cited, while the overall number of works tweeted by the user was found to have a negative correlation.

| | | Predicted | |
|---|---|---|---|
| | | False | True |
| **Cited** | False | 78,038 | 7619 |
| | True | 5204 | 9139 |

Table 4. Model predictions

Using this model to predict which tweeted works would be cited by the tweeter on a sample of 100,000 randomly selected tweets, with tweets used in training the model automatically excluded from the selection pool, resulted in an overall accuracy rate of 87.2%. As the overall rate of citations for tweeted works was 14.4%, this represents only a slight improvement in accuracy over a trivial prediction that no tweeted publications are cited. Recall rate of actual cited works/tweets for the model was 63.7%, while precision of predicted citations was 54.5%.

## 5. Discussion

The results of our study indicate that various geographic, socio-topical, and individual dimensions relating to an author and a tweeted publication influence the likelihood of a tweeted work being cited. Tweeters are more likely to cite works that are affiliated with their same institution, possibly indicating greater topical similarity or relevance, or possible intellectual involvement with colleagues within their own institution. In this way, cited tweets are influenced by geographic proximity and privy, perhaps, to the institutional dynamics of scholarship. This finding aligns with our socio-topical results, which show that tweeters are more likely to cite work they (co-)author due to topical proximity; and these collaborations are more likely to occur with colleagues within their same institution or country. This may result from an aim to increase the social capital of individual institutions or nations and subsequently contribute to this outcome, as well as augment the impact of work produced within the same linguistic and cultural contexts. This also sheds light on the heterogenic uses of social media; scholars citing their own work may indicate how Twitter can be used as a platform for increasing the visibility of one's own scholarship, establishing oneself as an expert in a domain, or extending one's social capital (Haustein, 2016; Haustein et al., 2016).

Moreover, our results (particularly the logistic regressions model), demonstrate how the topical similarity of a tweeted work to one's own research and field of study is highly influential on the relationship between the tweet and its eventual citation, confirming findings by Mongeon et al. (2018). That subfield has greater influence than fields or domains shows that tweeters are citing publications specifically relevant to their work, and less if they only relate in a more general sense to their disciplinary area. This is again exemplified by tweeted publications with no link exceeding the citation rate of those with links to domain and field, displaying only peripheral connection with little relation to overall topical relevance. Tweeters are also more likely to cite works published in journals in which they too have published, demonstrating the disciplinary circles that influence how scholars interact with research, and reifying the importance of topical similarity in the relationship between tweets and citations. The interdisciplinarity of certain disciplines may result in variance in citation distance from tweets while others may gravitate closely around a select few publications.

Finally, individual dimensions depicted in our results illuminate how academic age and the characteristics of a tweeter's scholarly career (total number of tweeted works, published works, and distinct references) influence their citation activity of the publications they tweeted. The plateau of citation rates depicted in the total number of published works and distinct references aligns with the negative trend shown for later-career researchers in Figure 3 depicting academic age. As careers progress and researchers publish more, they are less likely to cite what they tweet. This may indicate that researchers may be more active on Twitter at the start of their careers and aim to make their work and scholarly presence more visible to their peers, and later in their career they are less likely to engage with work on social media, correlating with a drop in their likelihood to cite tweeted works. Interestingly, the more works researchers tweet, the less likely they are to cite them, potentially indicating that less frequent tweeters are more selective in the works they choose to disseminate on social media. Those that tweet a great deal may instead focus on those that are more relevant for them from a citation point of view or engage with work on Twitter for a diverse range of reasons not always in relation to their own work and future citations, substantiating Bowman et al.'s (2015) contention that researchers tweet for both professional and personal purposes, and that motivations for citing and tweeting academic works by these more active tweeters do not necessarily align.

## 5.1. Limitations

This study has several limitations. First, it only considers Twitter counts and does not analyze other forms of altmetrics. The Open Dataset of Scholars on Twitter used to match Twitter users with researchers is a limited dataset of authors with at least one publication. Our dataset created with Crossref Event Data only considers works with DOIs. Additionally, errors with OpenAlex disambiguation may incorrectly attribute authors to publications.[5] Further, this study does not consider the influence of time and sequences of events on the correlation between tweets and citations. Finally, by gathering OpenAlex data from a May 2022 data dump, citations accumulated past that period are excluded from our data. We acknowledge that our study does not take into account disciplinary differences in citation practices; we did not expect that this consideration would substantially change our results and chose instead to focus our analysis on socio-topical characteristics of relationships between authors. Analyzing differences among disciplines may present a useful approach for future analyses.

## 6. Conclusion

As the use of altmetrics develops, understanding the relationship between altmetric activity and Twitter users is necessary for their meaningful interpretation. This study's analysis of over 5 million unique tweets reveals the geographic, socio-topical, and individual characteristics that influence the likelihood of researchers citing what they tweet. Our findings validate our hypothesis that topical, as well as social and geographic proximity (which overlap with topical proximity), positively increases the likelihood of citations and shows that topical similarity and geographic proximity bear significant influence on correlations between tweets and citations. Findings also affirm our hypothesis that the number of works and distinct references made by researchers will affect future citations positively.

---

[5] Author disambiguation in OpenAlex has changed somewhat since the May 2022 snapshot used in this project was obtained (OpenAlex [@OpenAlex_org], 2023). Details on the specifics of the disambiguation algorithm remain unpublished (Meyer, 2023), so we are unable to confirm how these changes would affect our results.

These findings demonstrate how individual characteristics of researchers on Twitter are important dimensions to consider when interpreting Twitter metrics around scholarly publications. Our findings have implications extending beyond tweeter behaviour; they elicit deeper consideration of the true meaning of altmetric activity, shifting attention from tweets as units of analysis to the researchers engaging with work in both social and scholarly realms, and the work itself. The relationship of the social media platform Twitter with scholarly communication can therefore be better understood by examining multiple dimensions (geographical, socio-topical, individual characteristics) associated with the characteristics of the actor, the work acted upon, and the relationships that exist between them outside of the social act itself.

## 6.1. Further research

Further research that aims to contextualize relationships between altmetric events and citations may wish to broaden the scope of an altmetrics analysis by bringing in other forms of altmetric data; discussions that aim to compare different social media metrics could use a similar approach which considers geographic, socio-topical, and individual dimensions of altmetric activities. Emerging altmetric data sources such as Mastodon could provide insights on the migration of researchers to new venues for the purposes of disseminating knowledge. Other individual-level features of the researchers like gender, country of origin, thematic specialization, reputation, etc. may also provide additional perspectives on how individual researchers are engaging on Twitter disseminating science. Additionally, other characteristics not included in the individual dimensions analyzed in this paper could be considered, such as the differentiating between original tweets and retweets, or other engagement indicators such as likes, replies, bookmarks, etc. (Fang et al., 2022). Further studies might choose to consider journal impact factor or highly cited publications to analyze socio-topical dimensions from an impact perspective, building off of Didegah et al.'s (2018) work. Content-level analysis of tweets could also be performed to better understand the direct causal aspects of a Tweeter's decision to engage with a work, shedding light on whether a work was tweeted for purposes of promotion, sharing, criticism, or other reasons. Disciplinary characteristics could also be investigated in more detail to determine if certain disciplines have higher or lower rates of citations. Further, authorship order could be an enlightening aspect of future analyses, illuminating whether tweeters are more likely to cite works in which they are first author, and how academic age may intersect with these elements.

## 7. Acknowledgements

The authors would like to thank Mercy Chikezie for her help with the literature review and the database design. Rodrigo Costas is partially funded by the South African DSI-NRF Centre of Excellence in Scientometrics and Science, Technology and Innovation Policy (SciSTIP).

## 8. Data availability

The dataset analyzed in this paper uses the Open Dataset of Scholars on Twitter created by Philippe Mongeon, Timothy Bowman, and Rodrigo Costas. This is a dataset of paired OpenAlex author_ids (https://docs.openalex.org/about-the-data/author) and tweeter_id.

The dataset includes 492,124 unique author_ids and 423,920 unique tweeter_ids forming 498,672 unique author-tweeter pairs. It is available on Zenodo at the following URL:

https://zenodo.org/record/7013518#.ZDlmpHZKi5c and the following article provides details about the matching process and links to R scripts: https://doi.org/10.1162/qss_a_00250

The dataset and R scripts produced for this analysis are available on Zenodo: https://doi.org/10.5281/zenodo.8039458

# 9. References

Adie, E. (2013, September 18). Gaming altmetrics. *Altmetric*.
https://www.altmetric.com/blog/gaming-altmetrics/

Alperin, J. P. (2015). Geographic variation in social media metrics: An analysis of Latin American journal articles. *Aslib Journal of Information Management*, *67*(3), 289–304.
https://doi.org/10.1108/AJIM-12-2014-0176

Archambault, E., Beauchesne, O. H., & Caruso, J. (2011). Towards a multilingual, comprehensive and open scientific journal ontology. *Proceedings of the 13th International Conference of the International Society for Scientometrics and Informetrics (ISSI)*, 66–77.

Bonetta, L. (2009). Should You Be Tweeting? *Cell*, *139*(3), 452–453.
https://doi.org/10.1016/j.cell.2009.10.017

Bornmann, L. (2015a). Alternative metrics in scientometrics: A meta-analysis of research into three altmetrics. *SCIENTOMETRICS*, *103*(3), 1123–1144.
https://doi.org/10.1007/s11192-015-1565-y

Bornmann, L. (2015b). Usefulness of altmetrics for measuring the broader impact of research: A case study using data from PLOS and F1000Prime. *Aslib Journal of Information Management*, *67*(3), 305–319. https://doi.org/10.1108/AJIM-09-2014-0115

Bornmann, L. (2016). What do altmetrics counts mean? A plea for content analyses. *Journal of the Association for Information Science and Technology*, *67*(4), 1016–1017. https://doi.org/10.1002/asi.23633

Bornmann, L., & Haunschild, R. (2018). Do altmetrics correlate with the quality of papers? A large-scale empirical study based on F1000Prime data. *PLOS ONE*, *13*(5), e0197133. https://doi.org/10.1371/journal.pone.0197133

Bowman, T. D. (2015). Differences in personal and professional tweets of scholars. *Aslib Journal of Information Management*, *67*(3).

Costas, R., de Rijcke, S., & Marres, N. (2021). "Heterogeneous couplings": Operationalizing network perspectives to study science-society interactions through social media metrics. *Journal of the Association for Information Science and Technology*, *72*(5), 595–610. https://doi.org/10.1002/asi.24427

Costas, R., Mongeon, P., Ferreira, M. R., van Honk, J., & Franssen, T. (2020). Large-scale identification and characterization of scholars on Twitter. *Quantitative Science Studies*, *1*(2), 771–791. https://doi.org/10.1162/qss_a_00047

Costas, R., Zahedi, Z., & Wouters, P. (2014). Do "altmetrics" correlate with citations? Extensive comparison of altmetric indicators with citations from a multidisciplinary perspective.

*Journal of the Association for Information Science and Technology*, n/a-n/a. https://doi.org/10.1002/asi.23309

de Winter, J. C. F. (2015). The relationship between tweets, citations, and article views for PLOS ONE articles. *SCIENTOMETRICS*, *102*(2), 1773–1779. https://doi.org/10.1007/s11192-014-1445-x

Dehdarirad, T. (2020). Could early tweet counts predict later citation counts? A gender study in Life Sciences and Biomedicine (2014–2016). *PLOS ONE*, *15*(11), e0241723. https://doi.org/10.1371/journal.pone.0241723

Díaz-Faes, A. A., Bowman, T. D., & Costas, R. (2019). Towards a second generation of 'social media metrics': Characterizing Twitter communities of attention around science. *PLOS ONE*, *14*(5), e0216408. https://doi.org/10.1371/journal.pone.0216408

Didegah, F., Bowman, T. D., & Holmberg, K. (2018). On the differences between citations and altmetrics: An investigation of factors driving altmetrics versus citations for finnish articles: JOURNAL OF THE ASSOCIATION FOR INFORMATION SCIENCE AND TECHNOLOGY. *Journal of the Association for Information Science and Technology*, *69*(6), 832–843. https://doi.org/10.1002/asi.23934

Eysenbach, G. (2011). Can Tweets Predict Citations? Metrics of Social Impact Based on Twitter and Correlation with Traditional Metrics of Scientific Impact. *Journal of Medical Internet Research*, *13*(4), e123. https://doi.org/10.2196/jmir.2012

Fang, Z., Costas, R., & Wouters, P. (2022). User engagement with scholarly tweets of scientific papers: A large-scale and cross-disciplinary analysis. *Scientometrics*, *127*(8), 4523–4546. https://doi.org/10.1007/s11192-022-04468-6

Ferreira, M. R., Mongeon, P., & Costas, R. (2021). Large-Scale Comparison of Authorship, Citations, and Tweets of Web of Science Authors. *Journal of Altmetrics*, *4*(1), Article 1. https://doi.org/10.29024/joa.38

Friedrich, N., Bowman, T. D., Stock, W. G., & Haustein, S. (2015). Adapting sentiment analysis for tweets linking to scientific papers. *Proceedings of the 15th International Society of Scientometrics and Informetrics Conference*, 107–108.

Hassan, S.-U., Saleem, A., Soroya, S. H., Safder, I., Iqbal, S., Jamil, S., Bukhari, F., Aljohani, N. R., & Nawaz, R. (2021). Sentiment analysis of tweets through Altmetrics: A machine learning approach. *Journal of Information Science*, *47*(6), 712–726. https://doi.org/10.1177/0165551520930917

Haustein, S. (2016). Grand challenges in altmetrics: Heterogeneity, data quality and dependencies. *Scientometrics*, *108*(1), 413–423. https://doi.org/10.1007/s11192-016-1910-9

Haustein, S., Bowman, T. D., & Costas, R. (2016). Interpreting "altmetrics": Viewing acts on social media through the lens of citation and social theories. In *Theories of informetrics and scholarly communication* (pp. 372–406). De Gruyter Saur Berlin, Boston.

Haustein, S., Costas, R., & Larivière, V. (2015). Characterizing Social Media Metrics of

Scholarly Papers: The Effect of Document Properties and Collaboration Patterns. *PLOS

ONE*, *10*(3), e0120495. https://doi.org/10.1371/journal.pone.0120495

Haustein, S., Peters, I, Sugimoto, C. R., Thelwall, M., & Larivière, V. (2014). Tweeting

biomedicine: An analysis of tweets and citations in the biomedical literature. *Journal of the Association for Information Science and Technology*, *65*(4), 656–669.

https://doi.org/10.1002/asi.23101

Holmberg, K., Bowman, T. D., Haustein, S., & Peters, I. (2014). Astrophysicists' Conversational

Connections on Twitter. *PLOS ONE*, *9*(8), e106086.

https://doi.org/10.1371/journal.pone.0106086

Holmberg, K., & Thelwall, M. (2014). Disciplinary differences in Twitter scholarly

communication. *Scientometrics*, *101*(2), 1027–1042. https://doi.org/10.1007/s11192-014-

1229-3

Jordan, K., & Weller, M. (2018). Academics and Social Networking Sites: Benefits, Problems

and Tensions in Professional Engagement with Online Networking. *Journal of

Interactive Media in Education*, *2018*(1), 1. https://doi.org/10.5334/jime.448

Ke, Q., Ahn, Y.-Y., & Sugimoto, C. R. (2017). A systematic identification and analysis of scientists on Twitter. *PLoS One*, *12*(4), e0175368.

https://doi.org/10.1371/journal.pone.0175368

Konkiel, S. (2016). Altmetrics: Diversifying the understanding of influential scholarship.

*Palgrave Communications*, *2*(1), 16057. https://doi.org/10.1057/palcomms.2016.57

Meyer, C. (2023, March 17). *Author Disambiguation Details* [Google Group post]. OpenAlex

    Users. https://groups.google.com/g/openalex-users/c/T7XOzPyI_wU

Mongeon, P. (2018). Using social and topical distance to analyze information sharing on social

    media. *Proceedings of the Association for Information Science and Technology*, *55*(1),

    397–403. https://doi.org/10.1002/pra2.2018.14505501043

Mongeon, P., Bowman, T. D., & Costas, R. (2023). An open data set of scholars on Twitter.

    *Quantitative Science Studies*, 1–11. https://doi.org/10.1162/qss_a_00250

Mongeon, P., Xu, S., Bowman, T. D., & Costas, R. (2018). Tweeting Library and Information

    Science: A socio-topical distance analysis. *23rd International Conference on Science and*

    *Technology Indicators (STI 2018), September 12-14, 2018, Leiden, The Netherlands*.

Nuzzolese, A. G., Ciancarini, P., Gangemi, A., Peroni, S., Poggi, F., & Presutti, V. (2019). Do

    altmetrics work for assessing research quality? *Scientometrics*, *118*(2), 539–562.

    https://doi.org/10.1007/s11192-018-2988-z

OpenAlex [@OpenAlex_org]. (2023, January 30). *We're launching a massive improvement to*

    *our author disambiguation algorithm, which will merge 100M+ duplicate authors! It's a*

    *big rollout, so you may see the occasion hiccup in the API or the data today. More info*

    *soon!* [Tweet]. Twitter. https://twitter.com/OpenAlex_org/status/1620101734428471296

Ortega, J. L. (2018). Disciplinary differences of the impact of altmetric. *FEMS Microbiology*

    *Letters*, *365*(7). https://doi.org/10.1093/femsle/fny049

Paul-Hus, A., Sugimoto, C. R., Haustein, S., & Larivière, V. (2015). Is there a gender gap in social media metrics? *ISSI*, *2015*, 15th.

Priem, J., Piwowar, H., & Orr, R. (2022). *OpenAlex: A fully-open index of scholarly works, authors, venues, institutions, and concepts* (arXiv:2205.01833). arXiv. https://doi.org/10.48550/arXiv.2205.01833

Robinson-Garcia, N., van Leeuwen, T. N., & Ràfols, I. (2018). Using altmetrics for contextualised mapping of societal impact: From hits to networks. *Science and Public Policy*, *45*(6), 815–826. https://doi.org/10.1093/scipol/scy024

Shu, F., & Haustein, S. (2017). On the citation advantage of tweeted papers at the journal level. *Proceedings of the Association for Information Science and Technology*, *54*(1), 366–372. https://doi.org/10.1002/pra2.2017.14505401040

Shu, F., Lou, W., & Haustein, S. (2018). Can Twitter increase the visibility of Chinese publications? *SCIENTOMETRICS*, *116*(1), 505–519. https://doi.org/10.1007/s11192-018-2732-8

Singh, L. (2020). A Systematic Review of Higher Education Academics' Use of Microblogging for Professional Development: Case of Twitter. *Open Education Studies*, *2*(1), 66–81. https://doi.org/10.1515/edu-2020-0102

Sud, P., & Thelwall, M. (2014). Evaluating altmetrics. *SCIENTOMETRICS*, *98*(2), 1131–1143. https://doi.org/10.1007/s11192-013-1117-2

Sugimoto, C. R., Work, S., Larivière, V., & Haustein, S. (2017). Scholarly use of social media and altmetrics: A review of the literature. *Journal of the Association for Information Science and Technology*, *68*(9), 2037–2062. https://doi.org/10.1002/asi.23833

Thelwall, M., Haustein, S., Larivière, V., & Sugimoto, C. R. (2013). Do altmetrics work? Twitter and ten other social web services. *PloS One*, *8*(5), e64841–e64841. https://doi.org/10.1371/journal.pone.0064841

Vásárhelyi, O., Zakhlebin, I., Milojević, S., & Horvát, E.-Á. (2021). Gender inequities in the online dissemination of scholars' work. *Proceedings of the National Academy of Sciences*, *118*(39), e2102945118. https://doi.org/10.1073/pnas.2102945118

Veletsianos, G., & Kimmons, R. (2016). Scholars in an increasingly open and digital world: How do education professors and students use Twitter? *The Internet and Higher Education*, *30*, 1–10. https://doi.org/10.1016/j.iheduc.2016.02.002

Webb, S. (2016). Twitter use in physics conferences. *SCIENTOMETRICS*, *108*(3), 1267–1286. https://doi.org/10.1007/s11192-016-2031-1

Yu, H., Xiao, T., Xu, S., & Wang, Y. (2019). Who posts scientific tweets? An investigation into the productivity, locations, and identities of scientific tweeters. *Journal of Informetrics*, *13*(3), 841–855. https://doi.org/10.1016/j.joi.2019.08.001

Zhang, L., & Wang, J. (2018). Why highly cited articles are not highly tweeted? A biology case. *SCIENTOMETRICS*, *117*(1), 495–509. https://doi.org/10.1007/s11192-018-2876-6

Zhu, J. M., Pelullo, A. P., Hassan, S., Siderowf, L., Merchant, R. M., & Werner, R. M. (2019).

Gender Differences in Twitter Use and Influence Among Health Policy and Health

Services Researchers. *JAMA Internal Medicine*, *179*(12), 1726–1729.

https://doi.org/10.1001/jamainternmed.2019.4027